\documentclass[12pt,preprint]{aastex}

\newcommand{\vap}{$v_{Ap}$}
\newcommand{\vac}{$v_{Ac}$}
\newcommand{\vae}{$v_{Ae}$}

\begin{document}

\title{Transverse oscillations of flowing prominence threads observed with Hinode}
\author{J. Terradas\altaffilmark{1,2}, I. Arregui\altaffilmark{2}, R. Oliver\altaffilmark{2}, and J. L. Ballester\altaffilmark{2}}
\altaffiltext{1}{Centre for Plasma Astrophysics, Katholieke Universiteit Leuven, Celestijnenlaan 200B, B-3001 Leuven, Belgium}
\altaffiltext{2}{Departament de F\'\i sica, Universitat de les Illes Balears, 07122 Palma de Mallorca, Spain}
\email{jaume.terradas@uib.es, inigo.arregui@uib.es, \\ramon.oliver@uib.es, dfsjlb0@uib.es}

\begin{abstract}
Recent observations with the Hinode Solar Optical Telescope display an active region prominence whose fine threads oscillate in the vertical direction as they move along a path parallel to the photosphere. A seismological analysis of this event is carried out by taking advantage of the small radius of these structures compared to the total length of magnetic field lines, i.e. by using the thin tube approximation. This analysis reveals that the oscillatory period is only slightly modified by the existence of the flow and that the difference between the period of a flowing thread and a static one is below the error bars of these observations. Moreover, although it is not possible to obtain values of the physical parameters, a lower bound for the Alfv\'en speed (ranging between 120 km s$^{-1}$ and 350 km s$^{-1}$) is obtained for each of the threads. Such Alfv\'en speeds agree with the intense magnetic fields and large densities usually found in active region prominences.
\end{abstract}


\section{Introduction}
Solar prominences are cold ($T\sim 10^4$ K) plasma structures embedded in the much hotter solar corona. The prominence material is much denser than its surroundings and, since it stays suspended  in the corona longer than the free-fall time, some force must counteract the downward force of gravity. Such an action is thought to be of magnetic origin and one of the prevailing ideas is that the prominence plasma lies on a dip of the magnetic field lines and that the field line curvature provides with the required upward force \citep[e.g.][]{AH07}.
A common feature of many prominences is that when observed with sub-arcsec resolution a complicated internal structure manifests: prominences are composed of many thin, parallel threads embedded in material which does not emit or absorb in prominence spectral lines. The thread plasma often flows along the magnetic field at speeds comparable to or larger than the local sound speed \citep{LERWB05}.

Prominence temperatures can usually be well determined by spectroscopic means, but other parameters are not so well-known or change widely between different prominences. \citet{T-H95} and \citet{PV02} give a compilation of various density determinations in prominences and show that this parameter varies by at least two orders of magnitude (namely from $10^9$ cm$^{-3}$ to $10^{11}$ cm$^{-3}$).
Magnetic fields show the same dispersion of values and range from a few G to 20--30 G in quiescent prominences \citep[e.g.][]{BLLS94,MTLC06,GHSSA07} and even higher values in active region prominences.

Given the difficulties in determining prominence parameters, prominence seismology has been put forward as an alternative method to probe the nature of these structures \citep[see][for some reviews of this topic]{OB02,BEOOS07}. The present work describes a seismological application to data obtained with the Hinode Solar Optical Telescope (SOT). 

\section{Observations}\label{observations}
The Hinode SOT observations analyzed in this work have been described in detail by \citet{Oetal07}. The data show an active region limb prominence composed of a myriad of thin horizontal threads that flow parallel to the photosphere. Some of these threads display vertical oscillations as they flow and these oscillations are synchronous, i.e. each thread moves entirely with a constant phase. The main features of the moving threads and their oscillations are given in Table~1 of \citet{Oetal07} and summarized in our table~\ref{table1}, so we just remark that these are thin structures whose width and length range from 360 to 660 km and from 1700 to 16,000 km, respectively. The period and the velocity amplitude of the vertical oscillations are in the range 135--250 s and 8--22 km s$^{-1}$, while the horizontal flow velocity varies between 15 and 46 km s$^{-1}$. A prominence thread is a cold plasma condensation that occupies a segment of a much longer magnetic tube. In the present case it is not easy to directly measure the length of such magnetic tubes, although \citet{Oetal07} estimate that the wavelength of the oscillations is at least 250,000 km and so the minimum length of the magnetic tubes is 125,000 km. Another estimation of this quantity can be obtained as follows: the oscillating threads lie parallel to the photosphere across the field of view of the SOT (which is 80,000 km wide), so that the full length of magnetic field lines is at least 80,000 km plus twice the vertical distance between the threads and the photosphere, which from table~\ref{table1} is at least 12,400 km. All this amounts to a minimum magnetic tube length of 100,000 km.

\section{Theoretical analysis of thread oscillations}\label{theory}
Since every thread displays a distinct period of oscillation it is reasonable to assume that its dynamics can be studied by isolating each of these structures from the others. Moreover, curvature and the non-linear terms in the MHD equations are neglected and the low-$\beta$ limit is used. Thus, it will suffice to consider the transverse vibrations of a straight magnetic tube embedded in the corona and with a flowing segment consisting of dense, prominence plasma.

\begin{figure}
\plotone{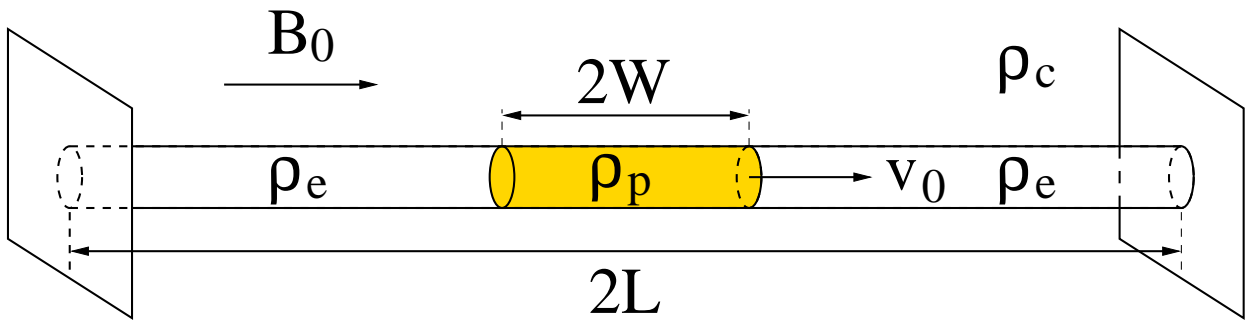}
\caption{Sketch of the magnetic and plasma configuration used to represent a flowing thread (shaded volume) in a thin magnetic tube. The two parallel planes at both ends of the cylinder represent the photosphere.}
\label{sketch_equilibrium}
\end{figure}

We then adopt an equilibrium model (Figure~\ref{sketch_equilibrium}) consisting of a cylinder of length $2L$ embedded in a uniform corona with density $\rho_c$. The plasma in the cylinder flows with speed $v_0$ and consists of a dense, middle section representing the observable thread in the Hinode SOT data, and an evacuated part of coronal material. These two regions have uniform densities $\rho_p$ and $\rho_e$. The plasma flow inside the magnetic tube results in the advection of the density structure with speed $v_0$.  We assume that at $t=0$ the thread occupies the center of the magnetic tube, so the density along the tube can be expressed as

\begin{equation}\label{density_flow}
\rho(z,t)=\left\{
\begin{array}{ll}
\rho_p & {\rm if }\;|z-v_0t|\leq W, \\
\rho_e & {\rm if }\;|z-v_0t|> W,
\end{array}
\right.
\end{equation}

\noindent where $2W$ is the length of the thread and the $z$-axis points along the tube, with $z=0$ the position of its middle point and $z=\pm L$ its end points. The magnetic field, $\bf B_0$, is the same inside and outside the cylinder and parallel to its axis.

The dynamics of the threads is studied taking into account that the ratio of the thread radius to the total length of field lines is a very small quantity, below $10^{-2}$ for all threads, so the thin tube approximation can be considered. A simple way to derive the equation governing the transverse oscillations is to follow the procedure of \citet{DR05}, who studied the normal modes of a structured magnetic cylinder, like the present one, for $v_0=0$. It is only necessary to include the two ingredients that are absent in that study: the flow inside the magnetic tube (which leads to the substitution of $\partial/\partial t$ by $\partial/\partial t+{\bf v}_0\cdot\nabla\equiv\partial/\partial t+v_0\partial/\partial z$ inside the cylinder) and the fully time-dependent situation (which prevents us from considering the time dependence $\exp(i\omega t)$).
It is worth to mention that the terms coming from the equilibrium flow can be ignored because, such as noted by \citet{DR05}, inside the cylinder the terms with derivatives along the tube, $\partial/\partial z$, are much smaller than those with radial or azimuthal derivatives. We then arrive to the remarkable result that the flow does not appear explicitly in the equations although it is present through the density; see equation~(\ref{density_flow}). Finally, a single differential equation that governs the transverse vibrations of the cylinder in the thin tube limit is obtained,

\begin{equation}\label{PDE}
\frac{\partial^2u}{\partial t^2}-c_k^2(z,t) \frac{\partial^2u}{\partial z^2}=0,
\end{equation}

\noindent where $u(z,t)$ is the transverse velocity component at the tube boundary, or in other words, $u(z,t)$ is the velocity component responsible for the observed lateral displacement of the threads. In this formula, which is a generalization of equation~(21) of \citet{DR05}, $c_k$ is analogous to the kink speed of a uniform magnetic tube and is given by $c_k^2(z,t)=2B_0^2/\mu[\rho(z,t)+\rho_e]$, where $\rho(z,t)$ must be substituted from equation~(\ref{density_flow}).

%

\subsection{Oscillations of steady threads}\label{steady_oscillations}
At first let us neglect the observed flows. The normal modes of a non-flowing prominence thread in the linear, low-$\beta$ limit have been studied by \citet{DOB02} and \citet{DR05}. Such structure supports Alfv\'en and fast waves. The first ones cause azimuthal (i.e. torsional) motions and so do not disturb the cylinder axis. Therefore, Alfv\'en modes are not the origin of the detected oscillations. Regarding fast waves, the only solution that gives rise to a displacement of the cylinder axis is the kink mode, hence it provides the most natural interpretation for the observed oscillations.

Of the two studies just mentioned, that of \citet{DR05} provides with the simplest dispersion relation because of the thin tube approximation.
The frequency of the kink mode is given by the smallest positive root of their equation (27), in which
%
%
%
$e=\rho_e/\rho_c$, $c=\rho_p/\rho_c$, $l=W/L$, and $\Omega=\omega L/$\vac, with $\omega$ the frequency and \vac\ the coronal Alfv\'en speed. The Alfv\'en speed in the thread and the evacuated parts of the tube are \vae\ $=e^{-1/2}$\vac\ and \vap\ $=c^{-1/2}$\vac, where $v_A^2=B_0^2/(\mu\rho_0)$ has been used. According to \citet{DR05} the kink mode frequency is practically constant for $0.6\leq e\leq 2$ (see their Figure~4). This result thus implies that, if the densities in the corona and in the evacuated part of the loop are similar, then $e=1$ can be taken without loss of generality. With this approximation equation~(27) of \citet{DR05} reduces to

\begin{equation}\label{dr_DR05mod}
\tan\left[\Omega(1-l)\right]-\sqrt{\frac{2}{1+c}}\cot\left[\Omega l\sqrt{\frac{1+c}{2}}\right]=0.
\end{equation}

The observations of \citet{Oetal07} provide us with two useful quantities: the oscillatory period (from which $\omega$ is computed) and the thread length 
(equal to $2W$ in our model). On the other hand, equation~(\ref{dr_DR05mod}) results in a single value of $\Omega$ once $c$ and $l$ are fixed and so it seems that this seismological problem has a unique solution. One must bear in mind, however, that the dimensionless variables $c$ and $l$ actually hide five dimensional variables ($\rho_c$, $\rho_p$, $L$, $W$, and \vac) and thus there are infinite solutions. Still some information can be extracted if we proceed as follows: let us start by selecting a thread from table~\ref{table1} and let us fix a total length of the magnetic tube, for which we have estimated a minimum value of 100,000 km, or $L\ge50,000$ km. Thus, $L$ and $W$ are known and so $l$ can be determined. Next, let us fix the density ratio, $c=\rho_p/\rho_c$, and let us calculate the smallest positive root, $\Omega$, of equation~(\ref{dr_DR05mod}). Now the definition of $\Omega$ can be used to determine the coronal Alfv\'en speed, \vac\ $=\omega L/\Omega=2\pi L/(\Omega P)$, where $P$ is the period, given in table~\ref{table1}. Finally, the Alfv\'en speed in the prominence is obtained, \vap\ $=c^{-1/2}$\vac. Hence, this procedure gives a unique pair of values \vac, \vap\ for each pair of $L$, $c$; in other words, if we impose a length of magnetic field lines and the density ratio, $c$, for a given thread, then the coronal and prominence Alfv\'en speeds are well determined. To display the results of this analysis we keep $L$ fixed and vary the density ratio, $c$, between a value slightly larger than 1 (extremely tenuous thread) and 1000 (very dense thread), and then plot \vap\ as a function of \vac\ for this range of $c$. Note that, although the coronal and thread densities cannot be computed from their respective values of the Alfv\'en speed because the magnetic field strength is unknown, the density of prominence plasmas is typically two orders of magnitude larger than that of the corona, which corresponds to $c$ of the order of 100.

The results for two of the six threads investigated by \citet{Oetal07} are plotted in Figure~\ref{fig_no_flow}; the other four threads have also been studied and lead to qualitatively similar results to those of this figure.
The leftmost point of the curves corresponds to $c\gtrsim1$, that is, to a thread density slightly larger than the coronal one (this point is not visible because it falls outside the range of the vertical axis). As a consequence, the solution \vap\ $\gtrsim$ \vac\ is obtained. When the density ratio is increased from this minimum value we move along each curve towards higher coronal Alfv\'en speeds. The left part of all curves displays a rapid variation of \vap, but for larger values of $c$ the Alfv\'en speed in the thread stabilizes at an approximately constant value, in spite of the unrealistically large coronal Alfv\'en speeds attained in these plots. From these results it is clear that a lower limit for the Alfv\'en speed in the threads can be established. For a length of magnetic field lines of 100,000 km this lower bound is between 200 km s$^{-1}$ and 350 km s$^{-1}$, with an exceptional lower value of 120 km s$^{-1}$ for thread \#6. These values of \vap\ are consistent either with a strong magnetic field, e.g. 50 G, and a large density, of the order of $1-8\times 10^{11}$ cm$^{-3}$, but also with a weaker magnetic field and lower density (10 G and $0.4-3\times 10^{10}$ cm$^{-3}$, for example).

\subsection{Oscillations of flowing threads}

Next the flowing motion of the threads is included in our model and again $\rho_e=\rho_c$ is considered for simplicity. Equation~(\ref{PDE}) has been integrated numerically using the code PDE2D \citep{Sewell05}. The large photospheric inertia in front of coronal perturbations is taken into account by imposing the boundary conditions $u=0$ at $z=\pm L$. The initial conditions for $u$ and $\partial u/\partial t$ must be carefully chosen in order to excite only the global transverse motions. This is accomplished by imposing the normal mode profile at $t=0$, given by equation~(24) of \citet{DR05} once the frequency is computed from equation~(\ref{dr_DR05mod}); this condition for $u(z,t=0)$ is accompanied by $\partial u/\partial t(z,t=0)=0$. The normal mode profile is more easily computed when the density is symmetric about the center of the tube, and this is the reason for choosing the thread initially placed in the tube center at $t=0$ in equation~(\ref{density_flow}). The duration of the numerical simulations is chosen so as to prevent the thread material from reaching the photosphere, and this typically corresponds to 4--6 kink mode periods. A uniform grid of 3000 points in the $z$-direction is used in all simulations, which ensures a minimum of 50 grid points in the thread.

\begin{figure}
\epsscale{0.8}
\plotone{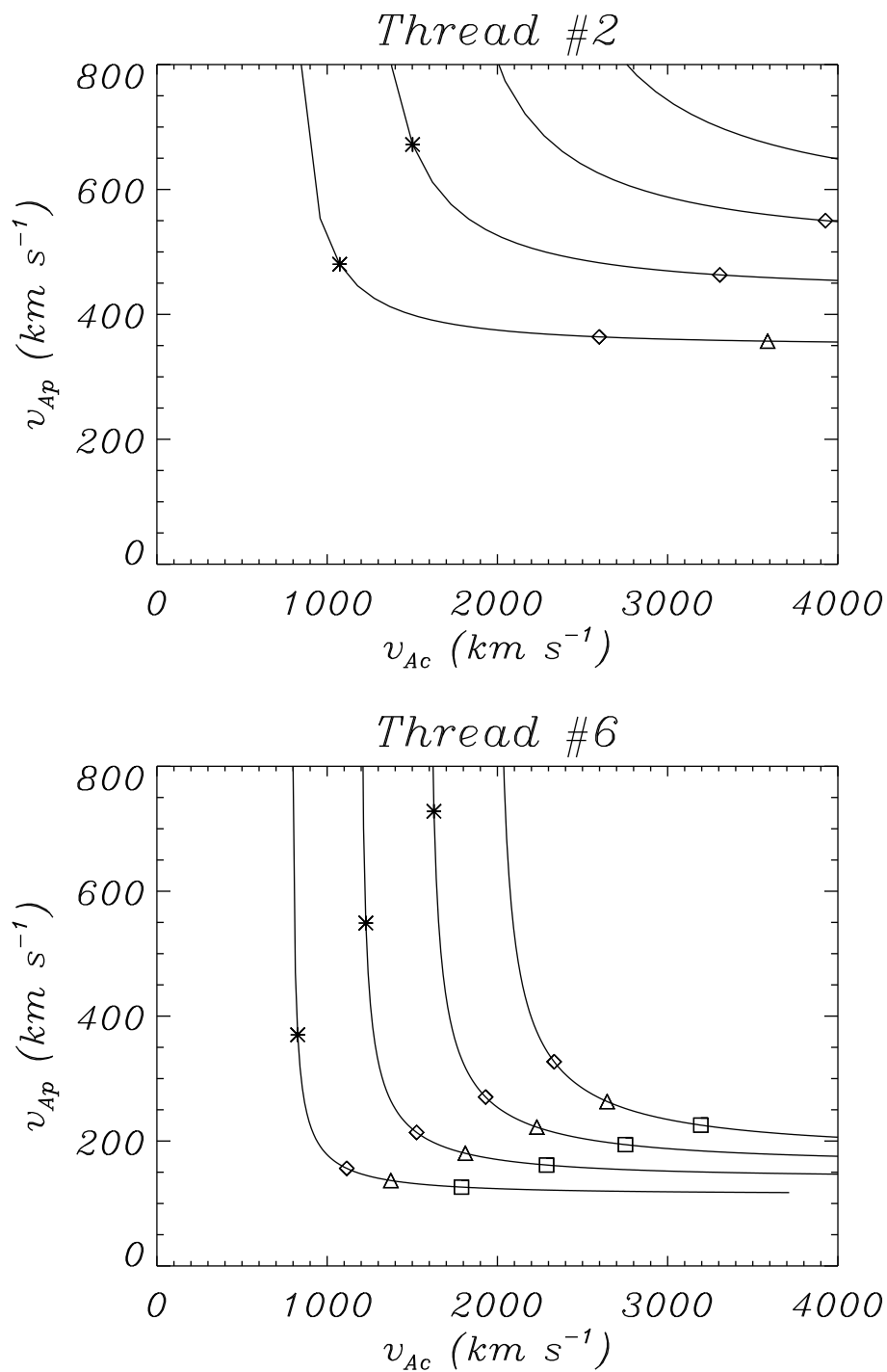}
\caption{Dependence of the Alfv\'en velocity in the thread (\vap) as a function of the coronal Alfv\'en velocity (\vac) for two of the threads studied by \citet{Oetal07}. In each panel, from bottom to top, the curves correspond to a length of magnetic field lines of 100,000 km, 150,000 km, 200,000 km, and 250,000 km, respectively. Asterisks, diamonds, triangles, and squares correspond to density ratios of the thread to the coronal gas $\rho_p/\rho_c\simeq 5,50,100,200$.}
\label{fig_no_flow}
\end{figure}

The numerical code has been run for each thread, whose length, flow speed, and oscillatory period are given in table~\ref{table1}, and for the parameter values (length of magnetic field lines, density ratio, and coronal and thread Alfv\'en speeds) corresponding to the symbols in Figure~\ref{fig_no_flow}. We first set $v_0=0$ and so expect to obtain the whole tube oscillating in the kink mode if the code works well. A representation of $u$ versus $z$ for different times shows that the whole thread oscillates in phase, which is consistent with the kink mode being excited by the initial perturbation. In addition, the power spectra of $u$ at various fixed positions present a strong power peak at a frequency that perfectly matches that of the kink mode. We are thus confident about the performance of our numerical code.

To assess the importance of the flow motion on the oscillatory period the above simulations are repeated with $v_0$ taken from table~\ref{table1}. Again it is found that the tube oscillates bodily in phase as it moves at the flow speed, $v_0$. The period is determined from the position of the peak in the power spectrum of the signal at a given point and then compared to the period given in table~\ref{table1}. In \S~\ref{steady_oscillations} the values in this table were used to obtain the results of Figure~\ref{fig_no_flow}, so any difference between the observed periods and those from the simulations with $v_0\neq 0$ comes from the thread flow. Our main conclusion here is that all the numerical simulations yield a period of transverse oscillations that is slightly smaller than that of the kink mode. The largest influence of the flow on the period is for short magnetic tubes and, for a given tube length, the largest deviations from the case $v_0=0$ occur for the largest density ratios, although it stops increasing for $c$ of the order of 100. For example, taking a total length of the magnetic tube of 100,000 km and $\rho_p/\rho_c=200$ this difference ranges between 3\% and 5\% for the six threads. For longer magnetic tubes this relative difference becomes even smaller. When these numbers are compared with the error bars of the period in table~\ref{table1}, whose minimum value is 6.8\%, it turns out that including the flow in the theoretical calculations leads to variations in the period that are undetectable with the cadence used to obtain the data.

To test the above results, obtained in the thin tube approximation, the linear, ideal MHD equations \citep[eqs.~(1a)--(1d) in][with the replacement of $\partial/\partial t$ by $\partial/\partial t+v_0\partial/\partial z$ inside the tube]{DR05} have been solved. The azimuthal dependence $\exp(i m\phi)$, with $m=1$, has been imposed and so a set of two-dimensional time-dependent equations has been investigated. No further approximations are made, i.e. the thin tube approximation is not used and the flow is maintained in the equations. The obtained results confirm the accuracy of the previous calculations.

\section{Discussion}\label{conclusions}

In this work we have performed a seismological analysis of some oscillating threads in an active region prominence observed with Hinode SOT. The horizontal threads flow along a path parallel to the photosphere and undergo simultaneous transverse oscillations that are interpreted as a signature of the kink mode of the whole magnetic tube in which the thread resides. Our numerical simulations prove that when the thread motion is included the global transverse oscillations persist and can be initiated by an external impulse. Therefore, the most natural explanation for these transverse thread oscillations is the excitation of the kink mode \citep[see also the discussion in][]{DNV08}, rather than Alfv\'en waves or motions along a helically twisted tube such as put forward by \citet{Oetal07}.

The available data are insufficient to derive well-constrained values of the physical variables, although it has been possible to establish a lower limit for the Alfv\'en speed in each of the threads. This lower bound comes from the assumption of the same minimum length for all threads, namely 100,000 km, but it could be larger if the actual length of the magnetic tube along which the thread flows is larger. To obtain other plasma parameters from the thread Alfv\'en speed one must make some assumptions. For example, if a magnetic field strength typical of active region prominences is used (50 G), then large densities ($1-8\times 10^{11}$ cm$^{-3}$) typical of active region prominences are obtained.

We have obtained a dispersion of the minimum Alfv\'en speed between 120 km s$^{-1}$ and 350 km s$^{-1}$ that can have multiple interpretations. On one hand, it is possible that not all threads have the same length, as has been assumed to derive these values, and that once the actual length is used all these values become the same. But on the other hand, one should consider the possibility of a highly inhomogeneous prominence, in which the magnetic field and density vary in space. This inevitably leads to an inhomogeneous distribution of the Alfv\'en velocity in the prominence, which is not surprising given the complex structuring of these objects.

Another important conclusion of this work is the insensitivity of the period to the flow velocity, which is a useless parameter in the present seismological problem. This has allowed us to simplify our study by considering a static situation. In addition, the thread thickness is another parameter of little importance given the much larger extent of the magnetic tube.

Some simplifications have been made to facilitate our theoretical investigation. First of all, field line curvature has been neglected since \citet{TOB06} have shown that it has little influence on the transverse vibrations of a flux tube, in the context of coronal loop oscillations \citep[see also][]{DDAP04}. In addition, the effect of the gas pressure in the MHD equations has been discarded since the plasma $\beta$ is much smaller than unity in the corona and in active region prominences. Another approximation has been to neglect non-linear terms in the MHD equations, which is justified since the velocity amplitude of transverse oscillations (whose largest value is 22 km s$^{-1}$; see table~\ref{table1}) is much smaller than the Alfv\'en speed both in the threads and in the corona.

A final comment can be made about the equilibrium model, that does not include gravity and therefore does not take into account the necessity of the magnetic structure to be able to support the dense prominence material. This effect has only been included in similar flux tube models by \citet{BP89,SD95,RSG99}, but the resulting structure is unstable. Hence, there is currently no prominence model that incorporates all relevant physics and leads to reasonable (stable) solutions and so using the simpler configuration studied in this paper is justified.


\acknowledgments
Funding provided under grants AYA2006-07637, of the Spanish Ministry of Education and Science, and PRIB-2004-10145 and PCTIB-2005GC3-03, of the Conselleria d'Economia, Hisenda i Innovaci\'o of the Government of the Balearic Islands, is also acknowledged.

\clearpage

\begin{deluxetable}{cccccc}
\tabletypesize{\scriptsize}
\tablecaption{ Summary of geometric and wave properties of vertically oscillating flowing threads analyzed by \citet{Oetal07}. $2W$ is the thread length, $v_0$ its horizontal flow velocity, $P$ the oscillatory period, $V$ the oscillatory velocity amplitude, and $H$ the height above the photosphere\label{table1}}
\tablewidth{0pt}
\tablehead{
\colhead{Thread} & \colhead{$2W$ (km)} & \colhead{$v_0$ (km s$^{-1}$)} & \colhead{$P$ (s)} & $V$ (km s$^{-1}$) & \colhead{$H$ (km)} 
}
\startdata
1 & 3600 & 39 & 174 $\pm$ 25 & 16 & 18,300 \\
2 & 16,000 & 15 & 240 $\pm$ 30 & 15 & 12,400 \\
3 & 6700 & 39 & 230 $\pm$ 87 & 12 & 14,700 \\
4 & 2200 & 46 & 180 $\pm$ 137 & 8 & 19,000 \\
5 & 3500 & 45 & 135 $\pm$ 21 & 9 & 14,300 \\
6 & 1700 & 25 & 250 $\pm$ 17 & 22 & 17,200
\enddata
\end{deluxetable}

\clearpage

\end{document}